# Less Stress, More Privacy: Stress Detection on Anonymized Speech of Air Traffic Controllers

Janaki Viswanathan*, Alexander Blatt*, Konrad Hagemann & Dietrich Klakow* (* Saarland University)

## Deutsche Zusammenfassung

Flugverkehrskontrolle (FVK) erfordert oftmals Multitasking unter Zeitdruck und ist dabei stets mit der Möglichkeit gravierender Konsequenzen im Falle von Fehlern verbunden. Eine solche Situation induziert mentale Beanspruchung mit Einfluss auf die menschliche Sprachproduktion. Die rechtzeitige Erkennung mentaler Beanspruchung und das Ergreifen angemessener Gegenmaßnahmen ist ein Schlüsselpunkt für die Aufrechterhaltung des hohen Sicherheitsniveaus im Luftverkehr. Die Verarbeitung von FVK-Sprachdaten ist mit engen rechtlichen Beschränkungen verbunden, wie z. B. der Datenschutz-Grundverordnung (DSGVO). Die Anonymisierung von Sprachdaten birgt eine Möglichkeit diese Einschränkungen zu adressieren. In diesem Artikel werden verschiedene Architekturen zur Stresserkennung für anonymisierte Sprache beim Fluglotsen evaluiert. Die besten Netzwerke erreichen eine Stresserkennungsgenauigkeit von 93,6% auf einer anonymisierten Version des Speech Under Simulated and Actual Stress (SUSAS)-Datensatzes und eine Genauigkeit von 80,1% auf einem anonymisierten FVK-Simulationsdatensatz. Dies zeigt, dass zumindest der Schutz der Persönlichkeitsrechte kein Hindernis für den Aufbau leistungsfähiger Deep-Learning-basierter Modelle für weiterführende Anwendungen basierend auf FVK Sprachdaten darstellen muss.

## Abstract

Air traffic control (ATC) demands multi-tasking under time pressure with high consequences of an error. This can induce stress. Detecting stress is a key point in maintaining the high safety standards of ATC. However, processing ATC voice data entails privacy restrictions, e.g. the General Data Protection Regulation (GDPR) law. Anonymizing the ATC voice data is one way to comply with these restrictions. In this paper, different architectures for stress detection for anonymized ATCO speech are evaluated. Our best networks reach a stress detection accuracy of 93.6% on an anonymized version of the Speech Under Simulated and Actual Stress (SUSAS) dataset and an accuracy of 80.1% on our anonymized ATC simulation dataset. This shows that privacy does not have to be an impediment in building well-performing deep-learning-based models.

*Keywords* - Air traffic control, stress detection, speech, privacy, anonymization

## Introduction

Air traffic controllers (ATCOs) constantly deal with a lot of information and need to choose the right procedure based on the circumstances and make quick decisions. The high level of responsibility along with the potentially fatal consequences of an error and working in shifts are known as prime sources of occupational stress [1].

Measures taken to prevent burn-outs and ATC-related incidents [2] include mandatory recovery breaks and continuous training of the ATCOs to handle stress and infrequent scenarios [1]. However, people cope with stress differently, which includes the behaviour during stress as well as the recovery time needed after stress. ATCO stress detection is an effective way to prevent incidents [3].

Monitoring ATCOs' mental state can be done in several ways. One approach is to use physiological measures like heart rate or respiration rate [4]. This has the drawback that these methods are intrusive and therefore not suitable for daily use in ATC. A less intrusive approach is to use operational speech data that are recorded anyway and are regularly deleted. A more direct approach to monitor the occurrence of stress is to use ATCO speech signals. Although stress detection for ATC speech is complicated by the fact that ATCOs are trained to remain calm even in stressful situations, Luig et al. [5] have already shown with simulated data that speech can be used to measure the workload of an ATCO. In their work, the authors argue that "stress" can be used as a term that describes "an individual's subjective capacity […] influenced by a multitude of factors" such as working conditions as well as "remarkable events and changes in private life" ([5], p.1-2). Single influences on this mental state are regarded as "stressors". By referring to the literature ([1], [22], [23], [24] and [25] cited after [5]) the authors describe stress as a factor that affects workload. According to Luig et al., the workload level is describing the subjective capacity utilization, which cannot be directly derived from the taskload level (related to the task complexity or size, e.g. traffic type or amount of traffic). Their approach has been to develop a speech analysis system for ATCO voice that indicates different factors of human stress with the goal to estimate from the stress level the ATCOs workload level. In contrary to that, in the work described in this article, subjective ISA workload measurements are used to estimate the stress levels, which in turn are used as a reference for a classification task based on ATC speech data.

A major restriction for any ATCO monitoring activities is privacy laws and regulations. Since ATC is a worldwide business, global and also local privacy laws must be met. With the rising collection of speech-assisted tools, there are also new guidelines that have to be met [6]. One way to avoid privacy-related issues is to remove personal information from the collected data. This can be done either on a text or speech level. On the text level, the entities can be masked or replaced that are linked to private information, for example, birth dates or phone numbers [7]. Since ATC speech is standardized and relies on a fixed phraseology[1], private entities are not as common as in normal speech. Therefore, it has been decided to focus on speech. On the speech level, anonymization assures that the original speakers - ATCOs or pilots, cannot be tracked back [8].

In the scope of this work, therefore, a stress recognition model for anonymized ATCO speech is proposed. In addition, a multiclass speaking style classification task is implemented to show that privacy does not have to be a barrier for speech processing.

## Related Work

Traditional speech-based stress or emotion identifying methods are rule-based or use Hidden Markov Models (HMMs) [9]. More recent approaches rely on deep learning methods. Tomba et al. [10] show that mean energy, mean intensity and mel frequency cepstral coefficients (MFCC) can be used to detect stress. Luig et al. [5] investigate different speech features for ATCO workload detection. They use the frequency of utterances spoken per minute as an indirect indicator of stress. Borghini et al. propose to measure ATCO stress directly from brain activities using methods like electroencephalography (EEG) [11]. In [12], the authors propose different model architectures based on deep-learning algorithms. They use convolutional layers to embed the relevant spectral input features and propose to add a long short-term memory (LSTM) network on top of the convolutional layers to capture the temporal components. The final multi-head attention layer can give more weight to the important parts of the input. This design is taken as the basis for our stress recognition model. Xu et al. [13] propose a similar architecture for emotion recognition and identify vocal tract length perturbation (VTLP) as a useful augmentation method for emotion recognition.

Speaker anonymization methods are benchmarked since 2020 in the Voice Privacy Challenge (VPC) [8]. For privacy evaluation, the VPC2020 considers various attack scenarios depending on the knowledge of the attacker. The first task is *unprotected* where both the users and the attackers use original data. The second task is *ignorant attacker* where the users anonymize their data but the attackers are unaware of it and use the original data. The third task is *lazy-informed* where both the users and attackers use anonymized data and the attacker also has access to the speaker identities. For the work at hand, the speaker anonymization method of Kai et al. [14] was used since it reaches equal error rates (EERs) above 40% on task II of the VPC2020 which indicates a high anonymization capability. The automatic speech recognition (ASR) method of the VPC2020 reaches a low word error rate (WER) of up to 10% on the anonymized speech which indicates that the anonymization of Kai et al. still allows the recognition of the spoken words.

However, other downstream tasks, i.e. applications of anonymized speech, are not investigated in the VPC2020. Therefore, an evaluation of emotion recognition has been included in this work.

## Experimental Setup

### Datasets

Our experiments are performed on the SUSAS [15] and DFS Munich approach simulation (DFS-MAS) datasets. The SUSAS dataset contains speech samples for different speaking styles. Nine speaking styles are considered - anger, fast, Lombard (increase of voice involuntarily when there is a background noise [21]), loud, clear, neutral, slow, soft, and question with 630 samples each except for neutral with 631 samples. Hence, the considered SUSAS dataset consists of 5671 samples in total. To enable binary stress detection, the following grouping of the speaking styles is suggested:

- STRESS: anger, fast, Lombard, loud
- NO-STRESS: clear, neutral, slow, soft

---

[1] ATC phraseology examples from the Federal Aviation Administration: https://www.faa.gov/air_traffic/publications/atpubs/aim_html/chap4_section_2.html

The label 'question' has been left out for our binary classification since it could occur in both stress and no-stress scenarios. Hence, there are 5041 samples for the stress detection task. An 80:20 split is done to create train and test sets and the train set is split again as 80:20 to create train and validation datasets. This is based on the approach used by Shin et al. [12] and results in a train | val | test split of 64%|16%|20%. This data split is used for all the experiments.

The DFS-MAS dataset was produced by the Deutsche Flugsicherung GmbH (DFS). It consists of ATC simulation data for Munich approach. The ATC speech samples were uttered by two male and two female ATCOs, with each having more than ten years of work experience. Following the approach by Luig et al. [5] described above, the workload level is used here as an approximation for the stress level of an ATCO. During the 90-minute simulation run, the workload of the ATCOs was measured every five minutes via an electronically presented pop-up questionnaire using the instantaneous self-assessment of workload technique (ISA) [16][17]. For binary stress detection on the DFS-MAS dataset, the stress labels are grouped according to the ISA workload labels:

- STRESS: high, excessive
- NO-STRESS: boring, relaxed, comfortable

The DFS-MAS dataset is highly imbalanced with 60 stress and 678 no-stress samples. Therefore, data augmentation methods such as VTLP and white noise addition are applied. To ensure that the distribution of the augmented data is the same across labels, the same number of augmented samples are generated for both the classes. The standard parameters in the 'nlpaug' package [26] are used. We generate ten different augmented versions of the stress samples - five using VTLP and five using white noise addition, while the majority class (No-Stress) is just augmented once per sample. Table 1 gives an overview of the data augmentation.

To test the performance of stress detection on anonymized data, an anonymized version of both the datasets is created. The anonymization method is described in the next section. The classification tasks are performed on both anonymized and non-anonymized data.

*Table 1: Summary of the augmented DFS-MAS dataset. The multiplication factors of the [train, validation, test] split represent the number of different copies created per clean sample.*

| Augmentation method | Stress [39, 9, 12] | No-stress [435, 108, 135] |
|---|---|---|
| None | [39, 9, 12] | [45, 18, 15] |
| VTLP | [39, 9, 12] * 5 | [195, 45, 60] * 1 |
| White noise | [39, 9, 12] * 5 | [195, 45, 60] * 1 |
| **Total** | **[429, 99, 132]** | **[435, 108, 135]** |

### Anonymization

As mentioned earlier, the lightweight voice anonymization (LVA) of Kai et al. [14] is used as the speaker anonymization method. Due to the high overall performance in the VPC2020 Tasks I, III, and V, waveform resampling is used as the anonymization method for the experiments here if not stated otherwise. Moreover, the gender-specific parameters are used for all the samples.

The resampling is based on the Waveform Similarity Overlap-Add (WSOLA) algorithm, which allows stretching the original speech signal by a factor $α$, while maintaining the correct pitch. Resampling this stretched signal by an $α$-times faster sampling frequency leads to the anonymized signal, which is of equal length as the original signal but varies, for example, in the pitch and formants.

### Speech Preprocessing

The ATC utterances are pre-processed before they are fed through the classification network. A Wiener filter [18] is applied first to remove noise. Furthermore, a pre-emphasis filter is applied which boosts the signal-to-noise ratio of the higher-frequency components since they are more susceptible to noise. Short-time Fourier transformation (STFT) is applied to generate the spectrogram. Then the log amplitude spectrogram is obtained by taking the logarithm of the amplitude component of the spectrogram. It is further converted to a mel spectrogram (MS) using the mel-frequency conversion formula [19] together with a filter bank of 128 filters. Two different speech representations are compared. The first one is obtained by applying the logarithm to the MS. This results in the log mel spectrogram (LMS) as network input. The second speech variant is generated by applying the discrete cosine transformation (DCT) to the LMS to generate the mel frequency cepstral coefficients (MFCC) [20]. Using MFCC has the advantage that the input data can be compressed without losing too much information by using the most informative DST coefficients and dropping the rest. For our experiments, 20 coefficients are used.

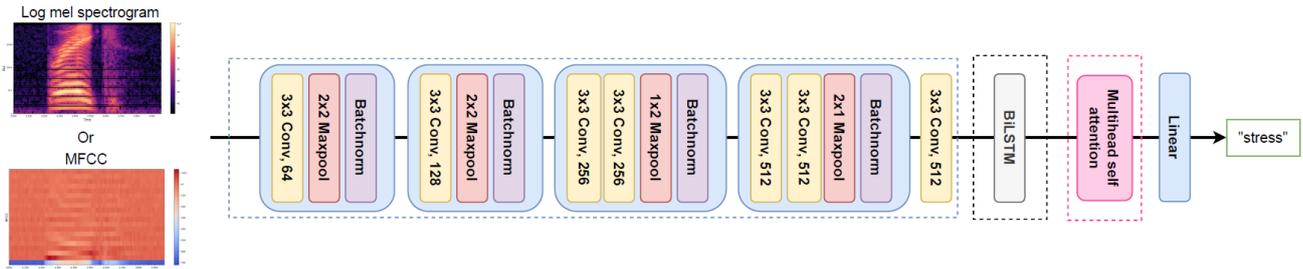

*Figure 1: Stress detection network depicting all the three architectures. The network was built incrementally. The blue dotted box represents the CNN, CNN along with the black dotted box represents CRNN, and CRNN. along with the pink dotted box represents CRNN+Attention model architecture.*

### Stress Detection Networks

Our stress detection networks are based on [12]. Three different architectures are investigated with increasing complexity - CNN, CRNN, and CRNN+Attention. They are built using different parts of the stack: CNN + LSTM + multi-head-Attention. Figure 1 shows the architecture of the models. Multi-head attention with four heads is used since this is the best-performing architecture in Shin et al. [12]. The experiments are repeated thrice and the mean and the standard deviation of the accuracies are calculated to check for robustness of the models.

## Results

### Architecture Comparison

The different architectures vary largely in their number of trainable parameters as shown in Table 2. This raises the question of whether the additional parameters lead to increased accuracy. The architecture comparison in Table 3 shows that either the CRNN or CRNN+Attention models have the highest accuracy for most of the experiments. The highest scores on the speaking style and stress classification tasks on the SUSAS dataset are reached by the CRNN architecture in combination with the LMS feature. This holds true for anonymized and non-anonymized data where the CRNN model outperforms the 11% larger CRNN+Attention model. In contrast, on the DFS-MAS dataset, the benefit of the additional attention layer of the CRNN+Attention model leads to a significant increase in accuracy of more than 5% in comparison with the CRNN model.

*Table 2: Comparison of architecture sizes for different speech representations.*

| Model architecture | Number of trainable parameters | |
|---|---|---|
|  | MFCC | LMS |
| CNN | 7,435,906 | 8,114,818 |
| CRNN | 9,012,866 | 9,691,778 |
| CRNN + Attention | 10,063,490 | 10,742,402 |

Replacing MFCC with LMS as input features leads to an average performance gain of 1-2%. This comes with the trade-off that the input dimension is increased by a factor of 6.4. Therefore, using MFCC as input is a valid alternative for devices with lower computational power.

### Stress Detection for ATC

ATC speech differs substantially from normal speech. It consists of a set of phraseologies that allow for handling different situations, like landing, take-off, and emergencies. In addition to that, ATCOs are supposed to give clear and calm instructions even under stressful situations. Furthermore, it is difficult to get a properly labelled, well-balanced dataset particular to the ATC scenario. This makes stress detection in this domain challenging.

Table 3 shows the difference in the accuracy of stress detection between the SUSAS and the DFS-MAS dataset. Due to the challenges mentioned above, the mean accuracy on the DFS-MAS dataset is about 20% lower. In contrast to the SUSAS data, the more complex CRNN+Attention model reaches the highest accuracies independent of the input features and the anonymization. This is another indicator of the difficulty level of the DFS-MAS dataset. Nevertheless, our best model reaches a performance of 80.1% on the DFS-MAS dataset.

## Anonymization Impact

Table 3 allows the comparison of the model performance of anonymized and non-anonymized datasets trained using different model architectures and different speech features. On the SUSAS dataset, the models trained on the anonymized version have a mean average accuracy that is 1-2% less than its non-anonymized counterpart. For the CRNN+Attention network, anonymization even leads to a performance increase. Figure 2 gives a more detailed insight into the classification accuracy for each class. For both anonymized and non-anonymized data, the CRNN model with LMS feature classifies the majority of the classes correctly, with an accuracy of over 70%. On both datasets, the model has problems distinguishing similar classes, such as clear, neutral, and slow from one another. The majority of the performance drop from non-anonymized to anonymized data is due to the misclassification of neutral speech, where the accuracy drops from 70% to 57%. For the other classes, the accuracy difference is 7% or less.

The anonymization method of the target data for stress detection might not always be known. Therefore, the question is, how would the performance decrease if the inference data is anonymized while the model is trained on non-anonymized data. Comparing the first row of Table 4 with the best-performing models Table 3 shows

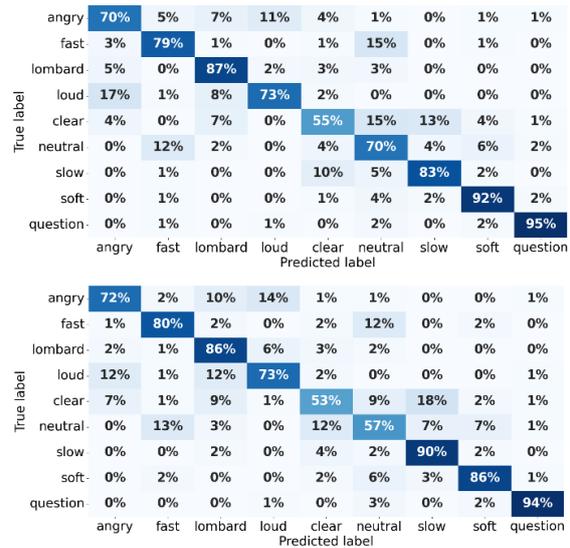

Figure 2: Confusion matrices of CRNN model with LMS feature on the non-anonymized (at the top) and anonymized (at the bottom) SUSAS dataset.

that the model performance decreases substantially for both model architectures.

While the accuracy of the CRNN+Attention model using MFCC drops from 93.9% to 80.1%, the accuracy of the CRNN model using LMS feature has an almost 19% drop - from 94.4% to 75.7%. The results are similar when the model is trained on raw SUSAS data and tested on anonymized SUSAS data.

Table 3: Mean accuracies of speaking style and stress recognition tasks on the SUSAS and DFS-MAS test sets. The standard deviation scores are given in brackets.

| Anonymized | Feature | Model architecture | SUSAS | | DFS-MAS |
|---|---|---|---|---|---|
| | | | 9 Speaking styles | Stress | Stress |
| No | MFCC | CNN | 75.6% [0.008] | 93.0% [0.009] | 74.2% [0.013] |
| | | CRNN | **75.8%** [0.008] | 93.1% [0.006] | 73.5% [0.012] |
| | | CRNN + Attention | 70.6% [0.029] | **93.9%** [0.006] | **75.6%** [0.039] |
| | Log Mel Spectrogram (LMS) | CNN | 76.8% [0.004] | 93.6% [0.002] | 66.9% [0.027] |
| | | CRNN | **77.7%** [0.006] | **94.4%** [0.004] | 66.9% [0.054] |
| | | CRNN + Attention | 73.9% [0.022] | 93.0% [0.005] | **71.6%** [0.042] |
| Yes | MFCC | CNN | **73.7%** [0.006] | 91.2% [0.002] | 71.8% [0.042] |
| | | CRNN | 72.3% [0.008] | 91.5% [0.005] | 69.5% [0.046] |
| | | CRNN + Attention | 71.5% [0.009] | **91.9%** [0.004] | **75.9%** [0.044] |
| | Log Mel Spectrogram (LMS) | CNN | 74.9% [0.008] | 92.5% [0.005] | 71.4% [0.006] |
| | | CRNN | **75.6%** [0.015] | **93.6%** [0.003] | 74.8% [0.036] |
| | | CRNN + Attention | 74.1% [0.003] | **93.6%** [0.002] | **80.1%** [3.384] |

On the ATC-relevant DFS-MAS dataset, the anonymization leads to an increase in performance. The best performing network, CRNN+Attention, trained and tested on anonymized data outperforms the best model for non-anonymized data by 4.5%. Since the nonaugmented DFS-MAS dataset is imbalanced with less than 100 stress utterances, the anonymization could act as an additional augmentation method. It should be noted that the CNN and CRNN models do not benefit from the anonymization, but they are also outperformed by the attention model by 1.4% to 9.7%.

*Table 4: Stress recognition cross-domain test accuracies. The best-performing models of Table 3 are used for testing. (A) represents the corresponding anonymized dataset.*

| Trained on | Tested on | MFCC | LMS |
|---|---|---|---|
| SUSAS | SUSAS (A) | 80.1% [ATTN] | 75.7% [CRNN] |
| SUSAS (A) | SUSAS | 80.6% [ATTN] | 78.7% [ATTN] |
| SUSAS | DFS-MAS | 50.2% [ATTN] | 50.2% [CRNN] |
| SUSAS (A) | DFS-MAS | 64.8% [ATTN] | 51.3% [ATTN] |
| SUSAS | DFS-MAS (A) | 56.2% [ATTN] | 45.3% [CRNN] |
| SUSAS (A) | DFS-MAS (A) | 72.3% [ATTN] | 52.1% [ATTN] |

## Cross-Domain Stress Detection

To the best of our knowledge, there are no publicly available stress-labelled ATC datasets. Therefore, it is also evaluated if it is possible to reach high stress recognition results on ATC data with a model that is trained on another domain. The results are shown in Table 4. For this, the best-performing SUSAS models, as marked in bold in Table 3, are used on the out-of-domain ATC data. In contrast to the results in Table 3 anonymizing the SUSAS dataset improves the cross-domain performance significantly by over 14% for the CRNN+Attention model with MFCC. The additional augmented data counteracts domain overfitting and leads therefore to a better generalization of the model. By adding anonymization also to the DFS-MAS test set, the performance increases over 22% in comparison with the non-anonymized datasets. With an accuracy of 72.3%, the difference to the best-performing model trained on the ATC data is below 8%. Interestingly, using MFCC as input gives consistently better cross-domain scores than using LMS as input. The higher information condensation in MFCC leads to a better generalization and hence avoids overfitting to the training domain, similar to anonymization.

## Conclusions

Our experiments show that anonymization is no obstacle to stress and speaking style recognition. In fact, it is observed that anonymization causes just a minor accuracy drop of 1-2% on the SUSAS dataset and even leads to a performance increase on the target ATCO speech of more than 4%. This probably comes down to the fact that anonymization can be seen as a data augmentation method, which could be beneficial, especially for low-resource tasks. Furthermore, we see that on the single speech style level, the performance drop is mainly due to the misclassification of neutral speech samples with, for example, similar clear speech samples. In other words, the classification results are stable through anonymization in the majority of the classes. In the cross-domain setting, it is shown that stress recognition models trained on out-of-domain data can be used to perform stress prediction on ATC. In this case, one should rely on MFCC as input since they generalize better than the LMS input. For our anonymization method, it is shown that if the anonymization method for ATC data is known, anonymizing the out-of-domain training data additionally improves the performance. Regarding the architectures, it is shown that a combination of MFCC and the CRNN model outperforms the CRNN+Attention models using the LMS feature in the speaking style recognition task, while having only 84% of its trainable parameters. This makes this model interesting if computational power is a limiting factor. Nevertheless, on the more demanding ATC data, the CRNN+Attention architecture outperforms the other networks by a margin, this holds also true for the cross-domain experiments.

For future work, we would like to explore different data augmentation methods which might increase the accuracy. Furthermore, we would like to investigate MFCC with different number of coefficients as an input feature since we observed equally good results as LMS. Another aspect to explore is transfer learning since it proved to be as good as the trained models on the DFS-MAS dataset. With transfer learning and the comparatively lower dimensional MFCC as an input feature, we could expand our work to have more practical applications where we could reduce the space and computational complexity to get live prediction and also train at the edge devices as and when we get new speech samples. By having a live stress detector, we could actively reduce the workload stress of ATCOs and avoid any incidents.

In summary, it is strongly suggested to test the incorporation of anonymization methods for privacy-critical tasks, especially for air traffic control.

*Acknowledgements: This work is based on the thesis "Identification of Stress from Anonymized ATCO Speech" to achieve a master's degree in Data Science and Artificial Intelligence at the Saarland University [27].*

## Abbreviations

| | |
|---|---|
| ASR | Automatic Speech Recognition |
| ATC | Air Traffic Control |
| ATCO | Air Traffic Controller |
| CNN | Convolutional Neural Network |
| CRNN | Convolutional Recurrent Neural Network |
| DCT | Discrete Cosine Transformation |
| DFS-MAS | DFS Munich Approach Simulation |
| DSGVO | Datenschutzgrundverordnung |
| EEG | Electroencephalogram |
| EER | Equal Error Rate |
| FVK | Flugverkehrskontrolle |
| HMM | Hidden Markov Models |
| LMS | Log Mel Spectrogram |
| LSTM | Long Short-Term Memory network |
| LVA | Light weight Voice Anonymization |
| MFCC | Mel Frequency Cepstral Coefficients |
| MS | Mel Spectrogram |
| STFT | Short-Time Fourier Transform |
| SUSAS | Speech Under Simulated and Actual Stress |
| VPC | Voice Privacy Challenge |
| VTLP | Vocal Track Length Perturbation |
| WER | Word Error Rate |
| WSOLA | Waveform Similarity Overlap-Add |